\documentclass{article}
\title{\bf A Potts/Ising Correspondence on Thin Graphs
}
\author{ {\it D.A. Johnston}\\
         Dept. of Mathematics\\
         Heriot-Watt University\\
         Riccarton\\
         Edinburgh, EH14 4AS, Scotland}
\date {20th April 1998}         
\begin{document}
  \maketitle
                      {\Large
                      \begin{abstract}
%
We note that it is
possible to construct a bond vertex model 
that displays $q$-state Potts criticality 
on  an ensemble of $\phi^3$ random graphs
of arbitrary topology, which we
denote as  ``thin'' random graphs in contrast to the fat graphs
of the planar diagram expansion.

Since the four vertex model in question
also serves to describe the critical behaviour of the Ising model in field,
the formulation reveals an isomorphism between the Potts 
and Ising models on thin random graphs. On planar graphs a similar correspondence
is present only for $q=1$, the value associated with percolation.

%
                        \end{abstract} }
%
  \thispagestyle{empty}
%
%
  \newpage
%
                  \pagenumbering{arabic}

\section{Introduction}

A description of the Potts model on planar random graphs as an
ice type vertex model on the associated medial graphs 
was developed some years ago by Baxter et.al.
\cite{1}. In this paper we formulate
a bond vertex description  of the Potts model on {\it non}-planar random graphs,
which we
call ``thin'' graphs to distinguish them from the fat graphs
which appear in a planar graph expansion. 
The study of spin and vertex models on such thin random graphs is of 
interest as a way of obtaining mean field theory results \cite{2a, 2b,2c,3} without the
boundary problems of the Bethe lattice or the inconvenience of infinite range
interactions. Mean field behaviour is present because the
thin graphs look locally like a Bethe lattice, but all the
branches of the tree-like Bethe lattice structure are closed by 
predominantly large loops. Planar graphs on the other hand have a 
loop distribution that is
strongly peaked on short loops and a fractal-like baby universe structure
\cite{4}.

Nonetheless,
the methods used for calculating the partition functions of spin models
on thin graphs may still be borrowed from the study of
the planar random graphs \cite{5}.
These methods are based on the observation that planar graphs can be thought of as
arising as Feynman diagrams in the perturbative expansion of
integrals over $N \times N$ Hermitian matrices. Each edge in such a Feynman diagram is ``fat''
or ribbon-like, since it carries two matrix indices.
In the $N \rightarrow \infty$  limit fat graphs of planar topology are picked out, since
there is an overall factor of $N^{\chi}$ for any graph, where $\chi$ is the Euler
characteristic. 
The 
$N \rightarrow 1$ limit, on the other hand,  weights all topologies equally. In this case
the matrix fat graph propagators degenerate to scalars, so we denote the
generic random graphs of the $N \rightarrow 1$ limit as ``thin'' graphs.
The Feynman diagram approach used in studying the statistical mechanics
of models on planar random graphs \footnote{Which is equivalent to coupling
the models to 2D quantum gravity.}
still applies for the thin graph case too. In fact, life is even easier in this case since
one is dealing with scalar rather than matrix integrals.

The partition function for a  statistical mechanical model on an ensemble
of thin random graphs with $2m$ vertices  can be defined
by means of an integral of the general form \cite{2a}
\begin{equation}
Z_m \times N_m = {1 \over 2 \pi i} \oint { d \lambda \over
\lambda^{2m + 1}} \int { {\prod_{i} d \phi_i  \over 2 \pi \sqrt{\det K}}
\exp (- S )},
\label{part}
\end{equation} 
where the contour integral over the vertex coupling $\lambda$ picks out
graphs with $2m$ vertices, $S$ is an appropriate action, $K$
is the inverse of the quadratic form in this action, 
and the $\phi_i$ are sufficient variables to describe the matter in the theory.
The factor $N_m$ gives the
number of undecorated  (without matter) graphs in the class of interest and disentangles this 
usually factorial growth
from any phase transitions. For the class of thin $\phi^3$ (3-regular) random graphs
we discuss here
\begin{equation}
N_m = \left( {1 \over 6} \right)^{2m} { ( 6 m - 1 ) !! \over ( 2 m ) !!
}.
\end{equation}
In the large $m$, thermodynamic, limit saddle point methods may be used
to evaluate equ.(\ref{part}). 
The saddle point equation for $\lambda$ may be decoupled
by scaling it out of the action as an overall factor, leaving any critical behaviour
residing in the saddle point equations for the matter fields $\phi_i$.
Phase transitions appear as an exchange of dominant saddle point.

\section{The Potts Model on Thin Graphs}

The Hamiltonian for a $q$-state Potts model can be written 
\begin{equation}
H =   \beta \sum_{<ij>}  \delta_{\sigma_i, \sigma_j} 
\end{equation}
where the spins $\sigma_i$ take on $q$ values.
The action which generates the correct Boltzmann weights 
for the $q$-state Potts model on $\phi^3$ graphs to be used in equ.(\ref{part}) is \cite{3,6}
\begin{equation}
S = { 1 \over 2 } \sum_{i=1}^{q} \phi_i^2 - c \sum_{i<j} \phi_i \phi_j -{\lambda \over 3} \sum_{i=1}^q \phi_i^3
\label{qstate}
\end{equation}
where $c$ is $1/ (\exp( 2 \beta ) + q - 2)$.
Since for any $q$ one finds a high temperature, disordered phase,  solution of the form $\phi_i= 1 - (q-1)c, \forall i$
bifurcating to a broken symmetry, ordered phase,  solution $\phi_2= \ldots \phi_{q-1} \ne \phi_1$
at $c=1/(2 q - 1)$ an effective action with only two variables $\phi, \tilde \phi$, where 
$\tilde \phi = \phi_1, \; \; \phi = \phi_2, \phi_3, \ldots , \phi_q$, suffices to describe the transition \cite{3}
\begin{eqnarray}
S_1 = {1 \over 2} ( q - 1) \left[ 1 - c ( q - 2) \right] \phi^2  - { \lambda \over 3} ( q -1) \phi^3
+ {1 \over 2} \tilde \phi^2  - {\lambda \over 3} \tilde \phi^3 - c ( q -1 ) \phi \tilde \phi.
\label{app1}
\end{eqnarray}
In the high temperature phase $\phi = \tilde \phi$ and this collapses to
\begin{eqnarray}
S_2 = { q \over 2} (1 - c ( q - 1) ) \phi^2 - { \lambda q \over 3} \phi^3.
\label{app2}
\end{eqnarray}

The bifurcation point is {\it not} the first order transition point displayed by the model
for $q>2$, but
rather a spinodal point. The first order transition is pinpointed by observing that
the free energy in the saddle point approximation is
the logarithm of the action $S$  
so the first order transition point is given by the $c$ value, and hence
temperature satisfying
$S_1 = S_2$.
This gives the critical value of $c$ as
\begin{equation}
c = { 1 - ( q - 1)^{-1/3} \over q -2},
\label{ccrit}
\end{equation}
in agreement with other mean field approaches \cite{7}.

The nature of the phase diagram can best be clarified by examining
a diagram of the magnetisation 
defined by
\begin{equation}
m = { \tilde \phi^3 \over \left( \tilde \phi^3 + (q-1) \phi^3 \right)}
\label{mag}
\end{equation}
$vs$ $c$, as plotted in Fig.1
for $q=4$. We can see that the bifurcation point at {\bf P}, $c = 1 / ( 2 q - 1)$ for general $q$,
lies below the first order transition point at {\bf Q},  $c = [ 1 - ( q - 1)^{-1/3} ]  / (  q -2 )$.
The second spinodal point at {\bf O}, 
$c = [ q -1 - 2 \sqrt{ q - 1}] / [ (q - 1 ) ( q - 5)] $,
is given by the vanishing of a square root in the saddle point solution.
As $q \rightarrow 2$ {\bf O, P, Q} coalesce,
the skewed pitchfork of Fig.1 becomes symmetrical and we recover the continuous mean field transition
of the Ising model -- which is equivalent to the $q=2$ Potts model. This is to be expected since
setting $q=2$ directly in the action of equ.(\ref{app1}) and the magnetisation of 
equ.(\ref{mag}) recovers the Ising action
and magnetisation.

Since $q$ appears explicitly as a parameter in equ.(\ref{app1}) the formalism is
ideally suited to studying
percolation on thin graphs as the $q \rightarrow 1$ limit of the Potts model, 
as well as the random resistor, $q \rightarrow 0$, 
and dilute spin glass, $q \rightarrow 1/2$, problems.

\section{A Potts Vertex Model}

In order to recast the Potts action as a bond vertex model
we first carry out the following rescaling on equ.(\ref{app1})
\begin{eqnarray}
\phi \rightarrow { 1 \over \sqrt{( q - 1) ( 1 - c ( q - 2) )}} \phi ,
\label{rescal1}
\end{eqnarray}
which gives both the quadratic terms the canonical coefficient of $1/2$,
followed by the linear transformation
$\phi \to (X - Y ) / \sqrt{2}$, $ \tilde \phi  \to ( X + Y ) / \sqrt{2}$
and a further rescaling of the quadratic terms and $\lambda$
to obtain
\begin{eqnarray}
S = \frac{1}{2} ( X^2 + Y^2) -  { \lambda \over 3} { ( 1 + v)  \over 2} \left[ X^3
+  3  \kappa^*  X Y^2 \right]  -  { \lambda ( \kappa^*) ^{3/2} \over 3} { ( 1 - v)  \over 2} \left[ Y^3
+  { 3 \over  \kappa^*}   Y X^2 \right],  
\label{ereal2}
\end{eqnarray}
where $\kappa^* = ( 1 - \kappa ) / ( 1 + \kappa) $ and
\begin{eqnarray}
v &=&   { 1 \over  (q-1)^{1/2} (1 - ( q - 2 ) c)^{3/2} } \nonumber \\
\kappa &=&  \sqrt{ c^2 ( q - 1) \over 1 - ( q - 2) c }.
\end{eqnarray}
The notation in equ.(\ref{ereal2})
has been chosen to facilitate comparison with the Ising model in
field below.

We can see from the above that the action of equ.(\ref{app1}) 
for the Potts model on thin $\phi^3$ graphs can be transformed via some
rescalings and a linear transformation of the variables into a 4-vertex model
on the $\phi^3$ graphs. What is more, the vertex model is symmetric since the weight
depends only on the number of $X$ and $Y$ bonds at a vertex \footnote{On random graphs the notion
of orientation is lost, so there are fewer
possibilities for defining independent vertex weights than on regular lattices. 
On planar random graphs one can still define a cyclic ordering of bonds round
a vertex consistently, but even this is lost on thin graphs.}. 
The different vertices in the model are shown in Fig.2.
Although the result of  a straightforward transformation,
equ.(\ref{ereal2}) entails a surprising consequence. The action
for the Ising model (i.e. the $q=2$ state Potts model) 
in field on thin graphs \footnote{And on planar graphs too, if
we take $X,Y$ to be matrices.} 
is given by \cite{7a}
\begin{equation}
S = Tr \left\{\frac{1}{2} ( X^2 + Y^2) - g X Y + \frac{\lambda}{3} \left[
e^h X^3 +
e^{-h} Y^3 \right],
\right\}
\label{ising}
\end{equation}
where $g = \exp ( - 2 \beta )$ and $h$ is the external field. 
Carrying out
the  transformations
\begin{eqnarray}
X &\rightarrow& ( X + Y ) / \sqrt{2} \nonumber \\
Y &\rightarrow& (X - Y) / \sqrt{2}
\label{ortho}
\end{eqnarray}
followed by the rescalings $X \rightarrow X / ( 1 - g )^{1/2}, \; Y
\rightarrow Y / ( 1 + g)^{1/2}$, $\lambda \rightarrow \sqrt{2} \lambda  (
1 - g)^{3/2}$ again gives a four-vertex model \cite{8} 
\begin{equation}
S = \frac{1}{2} ( X^2 + Y^2) -  {\lambda \cosh(h) \over 3 } \left[ X^3 +3 g^* X Y^2 \right] -
{\lambda \sinh ( h ) (g^*)^{3/2} \over 3}  \left[  Y^3 + { 3 \over g^*} X^2 Y \right] 
\label{ibond}
\end{equation}
where $g^* = ( 1 - g) / ( 1 + g)$

We thus find that the vertex model action for the $q$ state Potts model
and that for the Ising model in field are isomorphic under the identifications
\begin{eqnarray}
\tanh ( h) &=& { 1 - v \over 1 + v} \nonumber \\
g & = & \kappa
\end{eqnarray}
In fact, backtracking to equ.(\ref{app1}) we can see that the rescaling
of equ.(\ref{rescal1}) transforms the action into that of an Ising model 
in field, even before the transformation to a vertex model
\begin{equation}
S \rightarrow { 1 \over 2} ( \phi^2 + \tilde \phi^2 ) - \kappa
\phi \tilde \phi -{ \lambda v  \over 3 } \phi^3 
-{\lambda \over 3} \tilde \phi^3. 
\end{equation}

The equivalence between the Potts and Ising models is surprising 
from the physical point of view since
we know already that the Potts models display  a first order transition
for $q>2$ \cite{3,7}
whereas the Ising model displays a (mean-field) second order transition.
Things become clearer when we consider the mapping of the Potts values
for $c( q, \beta)$ at the critical and spinodal points onto the parameters
$\kappa, v$, which play the role of
temperature and field in the Ising model (or more precisely
$\exp ( - 2 \beta)$ and $\exp ( 2 h )$). 
The corresponding values of $\kappa$ for {\bf O,P,Q} are shown in Fig.3
and for $v$ in Fig.4.
We find that the first order transition point in the Potts models at {\bf Q}, where
$c = [ 1 - ( q - 1)^{-1/3}]  / ( q -2)$, maps onto 
\begin{eqnarray}
\kappa &=& \pm {( q - 1 ) ^{2/3} - ( q -1 )^{1/3} \over q - 2} \nonumber \\
v &=& 1
\end{eqnarray}
where the sign on the right hand side of the expression for $\kappa$ is
chosen to give a positive answer depending on whether
$q <> 2$.
Remarkably we see that this corresponds to a zero-field point
in the Ising model,
with $\kappa < \kappa_c = 1/3$, the Ising critical value. Since this
means that $\beta > \beta_{critical \; Ising}$, this point lies
on the zero field line separating the two possible spin
orientations in the ordered phase, so the first order temperature
driven transition of the Potts model is mapped onto the field
driven transition of the Ising model.
As $q \rightarrow 2$ from above or below, $\kappa \rightarrow 1/3$, 
the Ising critical value, and the transition becomes the continuous
mean-field Ising transition.

The spinodal point at {\bf P} where $c = 1 / ( 2 q - 1)$
maps onto
\begin{eqnarray}
\kappa &=&  \sqrt{ q - 1 \over ( 2 q - 1 ) ( q + 1) }\nonumber \\
v &=& { 2 q - 1 \over ( q -1 )^{1/2} ( q + 1)}                                                   
\end{eqnarray}
which tends towards the standard mean field Ising transition point in zero-field
as $q \rightarrow 2$.
Similarly the other spinodal point at 
at {\bf O},
$c = [ q -1 - 2 \sqrt{ q - 1}] / [ (q - 1 ) ( q - 5)] $, 
maps onto
\begin{eqnarray}
\kappa &=&   { ( q - 1)^{1/4} \over ( 2 \sqrt{ q - 1} + 1)^{1/2} ( \sqrt{q - 1} + 2)^{1/2}} \nonumber \\
v &=&   { (\sqrt{q-1} + 2)^{3/2} ( q - 1)^{1/4} \over ( 2 \sqrt{ q - 1} + 1 )^{3/2}  }
\end{eqnarray}
so we see that both these points are not, in general, zero field points in the Ising transcription except
at $q=2$.

A natural question to ask is what is so special about {\bf Q} from the Ising point of view. If we look at the
effect of the scaling of equ.(\ref{rescal1}) on the original Potts action of equ.(\ref{app1})
and demand a $c$ value which gives $v=1$ (i.e. zero field) we get precisely   
$c = [ 1 - ( q - 1)^{-1/3}]  / ( q -2)$. The first order transition point at $c($ {\bf Q} $)$ is thus the only
point which maps onto zero field in the Ising model for general $q$.
 
\section{Discussion}

The main conclusion of this paper may be simply stated: the Ising 
model in field and the Potts model
on thin $\phi^3$ random graphs may be mapped onto one another. This can be seen
either by mapping them both to a four vertex model, or by directly rescaling the Potts
action. The equivalence is at root due to the fact that the Potts action
of equ.(\ref{app1}) requires only two variables, just like
the Ising model, even though the Potts
spins have $q$ states. 
We have also seen that the first order nature of the Potts transitions for $q \ne 2$ and the
continuous transition of the Ising model are not in contradiction, since the Potts transition
maps onto the field driven transition of the Ising model. 
The first order Potts transition point lies on the zero field Ising locus
at $\beta> \beta_{critical \;  Ising}$ and moves to the continuous Ising transition point as
$q \rightarrow 2$.

One might enquire as to whether similar relations could exist on planar random graphs
where the scalars in actions such as equ.(\ref{qstate}) are replaced by
$N \times N$ Hermitian matrices. Unfortunately, a two
variable (in this case matrix variable) effective action such as that in
equ.(\ref{app1}) does not appear to exist in such a case.
However, one can arrive at a $3q + 1$ vertex model for planar graphs by using Kazakov's
approach \cite{6} of introducing an auxiliary matrix to decouple the Potts interactions.
This transforms the matrix action for the $q$ state Potts model
\begin{equation}
S = Tr \left\{ { 1 \over 2 } \sum_{i=1}^{q} \phi_i^2 - c \sum_{i<j} \phi_i \phi_j -{\lambda \over 3} \sum_{i=1}^q \phi_i^3
\right\}
\label{matqstate}
\end{equation}
to
\begin{equation} 
S = Tr \left\{ { 1 \over 2 }  X^2 + { 1 + c \over 2 } \sum_{i=1}^{q} \phi_i^2 - \sqrt{c} \sum_{i} X \phi_i  -{\lambda \over 3} \sum_{i=1}^q \phi_i^3 
\right\} 
\end{equation}
which after a shift in $\phi_i$ and a rescaling may be written as
\begin{equation}
S = Tr \left\{ { 1 \over 2 }  X^2 + 
\sum_{i=1}^{q} { 1 \over 2 } \phi_i^2 - { \lambda \over 3}  \sum_{i=1}^{q} \left( X^3  + 3 v^{1/2} X^2 \phi_i + 3 v X \phi_i^2 + v^{3/2} \phi_i^3 \right)
\right\}
\label{planar}
\end{equation}
where $v = \exp ( \beta ) - 1$. In general this will not be equivalent to an Ising model since we have too many matrices,
but when $q=1$, which is related to the problem of percolation, we do recover an Ising-like action.

It would be of some interest 
to see whether this transcription might shed some light on either
percolation or the Ising model in field on planar random graphs.  
It would also be interesting to explore the relation between the planar graph Potts vertex model of equ.(\ref{planar}) and the 
medial graph vertex model of Baxter et.al. \cite{1}

\section{Acknowledgements}

This work was partially supported by a Leverhulme Trust Research Fellowship
and a Royal Society of Edinburgh/SOEID Support Research Fellowship.

%

%
\clearpage \newpage
\begin{figure}[htb]
\vskip 20.0truecm
\includegraphics{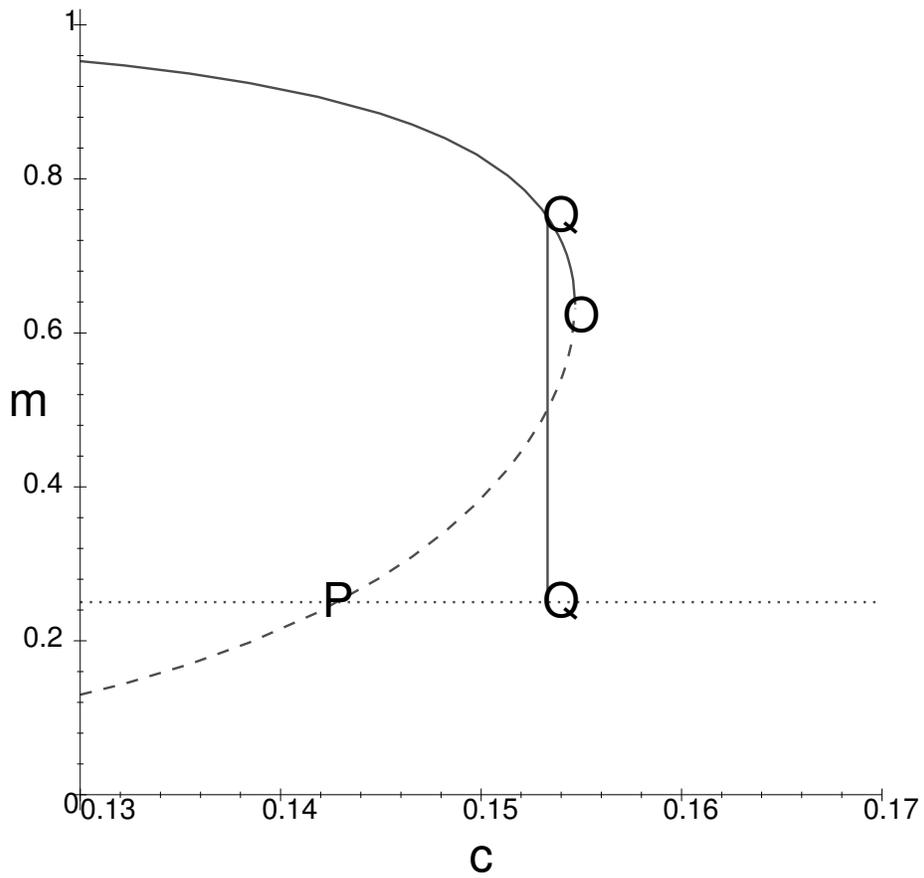}
\caption[]{\label{fig1} The magnetisation
$m$
for a $4$ state Potts model as calculated from
the saddle point solutions. The high temperature solution
is shown dotted, one low temperature solution solid and the other dashed.
The first order jump in $m$ is at {\bf Q}.}
  
\end{figure}
\clearpage \newpage
\begin{figure}[htb]
\vskip 20.0truecm
\includegraphics{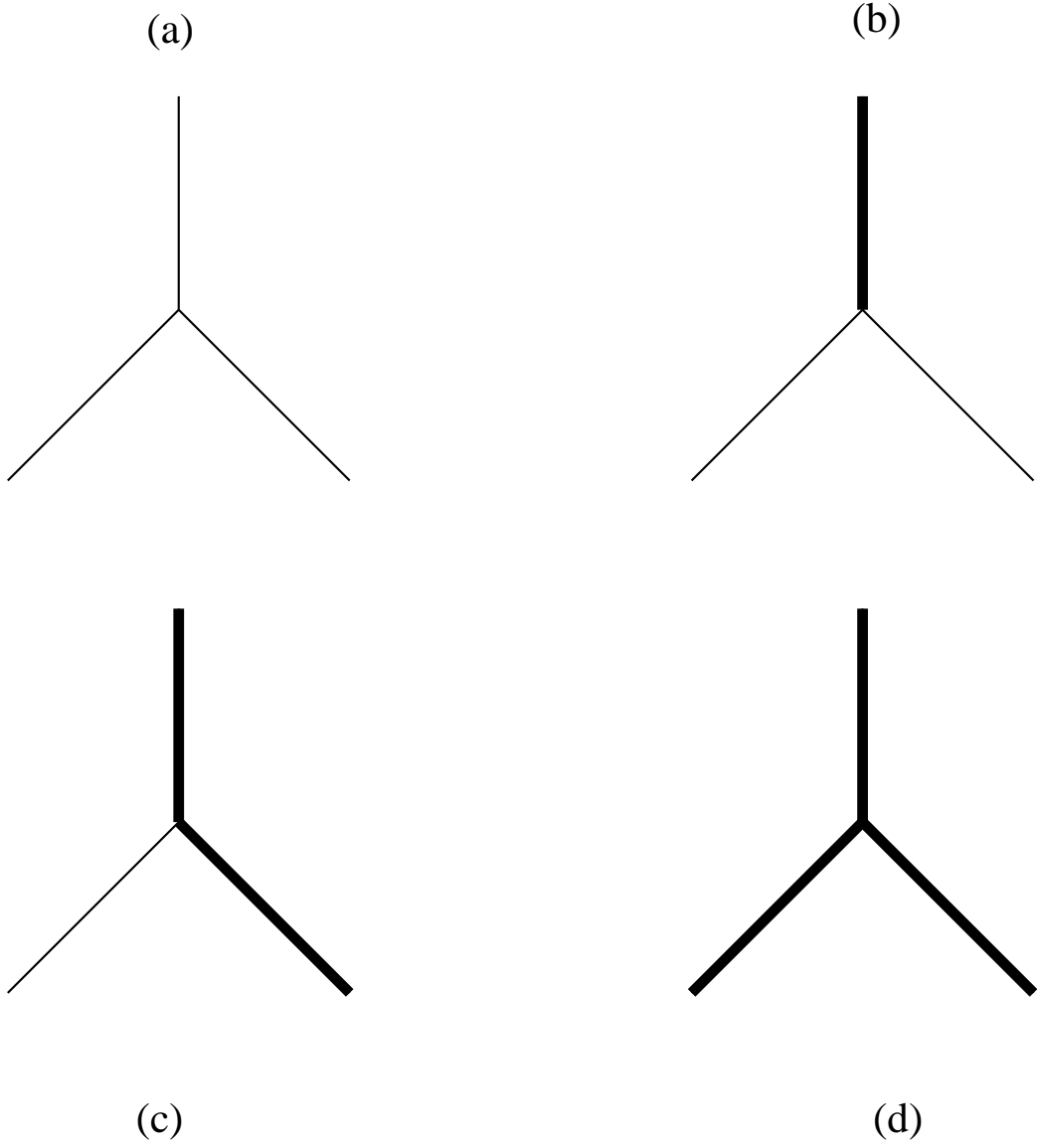}
\caption[]{\label{fig2} The possible bond vertices
which appear in the model.
The Potts weights on the random graph, which
can be read off from equ.(\ref{ereal2}), are:
$a = ( 1 + v ) / 2$  , $ b  = (\kappa^*)^{1/2} ( 1 - v ) /2$,
$c = \kappa^* ( 1 + v ) / 2$   and $d = (\kappa^*)^{3/2} ( 1 - v ) /2 $.}
\end{figure}
\clearpage \newpage
\begin{figure}[htb]
\vskip 20.0truecm
\includegraphics{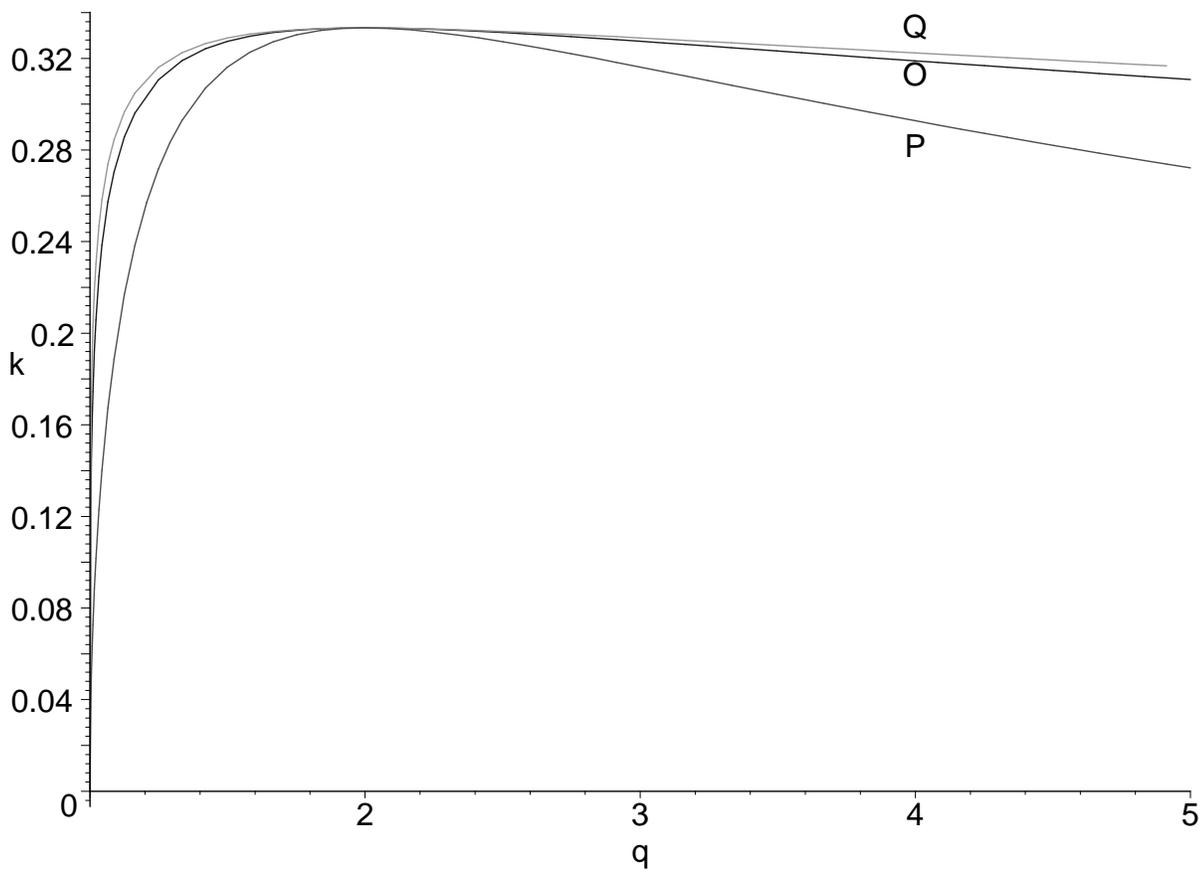}
\caption[]{\label{fig3} $\kappa$ {\it vs } $q$ for the
points {\bf O,P,Q}. The curves touch at $q=2$, $\kappa=1/3$ which
is the Ising critical point.}
\end{figure}
\clearpage \newpage
\begin{figure}[htb]
\vskip 20.0truecm
\includegraphics{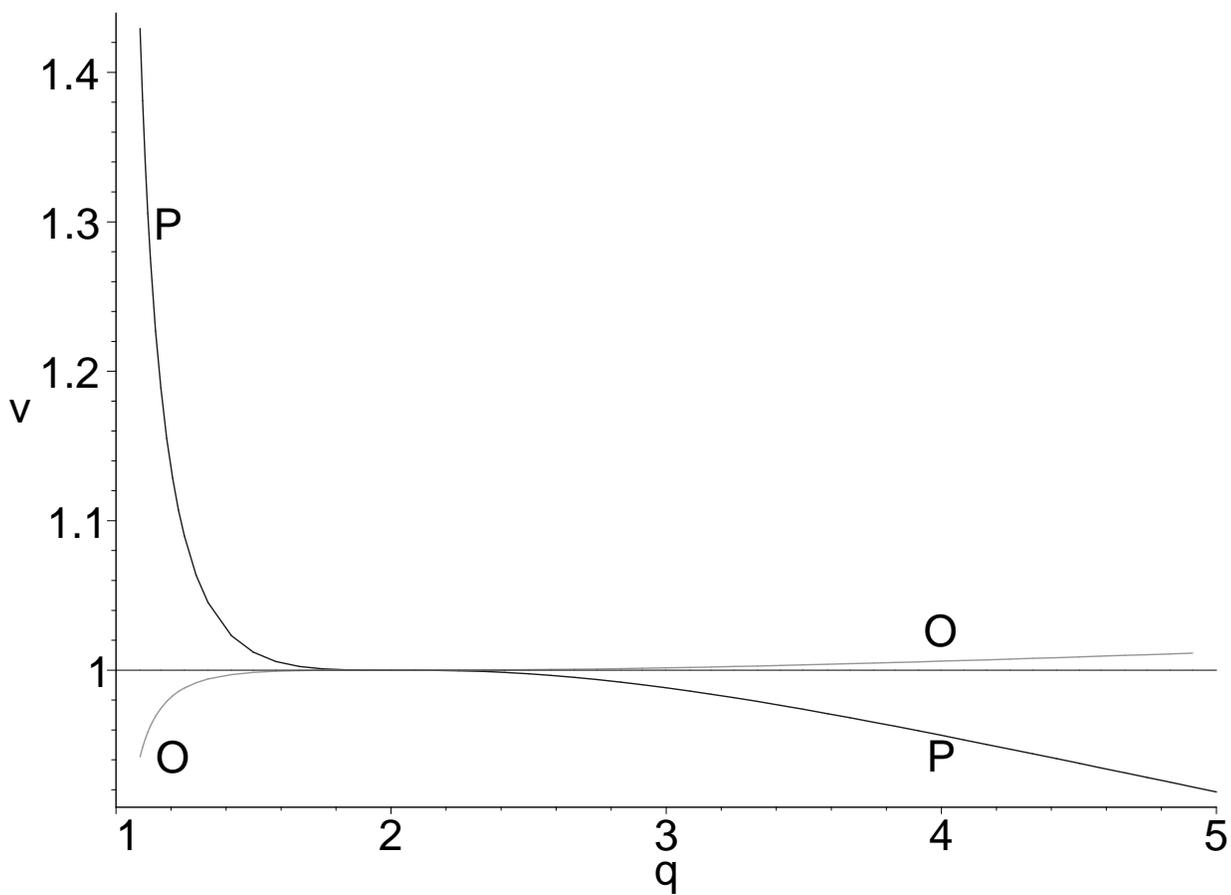}
\caption[]{\label{fig4} $v$ {\it vs } $q$ for the
points {\bf O,P,Q}. Since $v=1$ corresponds to zero-field
in the equivalent Ising model, we see that {\bf Q} is a zero-field point for
all $q$. For {\bf O, P} on the other hand $v=1$ only at $q=2$.}
\end{figure}
\end{document}